\documentclass[english]{article}
\usepackage[T1]{fontenc}
\usepackage[latin9]{inputenc}
\usepackage[letterpaper]{geometry}
\usepackage{graphicx}
\usepackage{esint}

\makeatletter

\DeclareRobustCommand{\greektext}{%
  \fontencoding{LGR}\selectfont\def\encodingdefault{LGR}}
\DeclareRobustCommand{\textgreek}[1]{\leavevmode{\greektext #1}}
\DeclareFontEncoding{LGR}{}{}

\providecommand{\tabularnewline}{\\}

\makeatother

\usepackage{babel}

\begin{document}

\title{TIME SERIES ANALYSIS OF THE RESPONSE OF MEASUREMENT INSTRUMENTS}

\author{Georgakaki D.$^{1}$, Mitsas Ch.$^{2}$, Polatoglou H.M.$^{1}$}

\maketitle
\begin{center}
$^{1}$Physics Department, Solid State Physics Section, Aristotle
University of Thessaloniki., 54124
\par\end{center}

\begin{center}
$^{2}$ Mechanical Measurements Dept., Hellenic Institute of Metrology,
Thessaloniki, 57022
\par\end{center}
\begin{abstract}
In this work the significance of treating a set of measurements as
a time series is being explored. Time Series Analysis (TSA) techniques,
part of the Exploratory Data Analysis (EDA) approach, can provide
much insight regarding the stochastic correlations that are induced
on the outcome of an experiment by the measurement system and can
provide criteria for the limited use of the classical variance in
metrology. Specifically, techniques such as the Lag Plots, Autocorrelation
Function, Power Spectral Density and Allan Variance are used to analyze
series of sequential measurements, collected at equal time intervals
from an electromechanical transducer. These techniques are used in
conjunction with power law models of stochastic noise in order to
characterize time or frequency regimes for which the usually assumed
white noise model is adequate for the description of the measurement
system response. However, through the detection of colored noise,
usually referred to as flicker noise, which is expected to appear
in almost all electronic devices, a lower threshold of measurement
uncertainty for this particular system is obtained and the white noise
model is no longer accurate.

\end{abstract}

\section{Introduction}

Exploratory Data Analysis (EDA) is an approach, contrast to classical
approach and Bayes theory, which allows the data itself to reveal
its underlying structure and model {[}1{]}. In particular, Time Series
Analysis (TSA), treats every experimental data set as a group of subsequent
observations in time \{$X_{t}$\} of a random variable X (measurand). 

This technique has been recently applied to the analysis of experimental
measurements collected in metrological laboratories in an attempt
to clarify the role of serially correlated data with stochastic characteristics
with regard to more realistic measurement uncertainty estimation (see
papers from Witt T.J. et al, Zhang N.F. et al). Via these analysis
methods it has been shown that if measurements are autocorrelated
then the usual \textquotedblleft{}classical variance\textquotedblright{}
as expressed through the standard deviation of the mean of a set of
repeated measurements can lead to an underestimation of the measurement
uncertainty {[}2,3{]}. An alternative approach for the variance calculation
of autocorrelated measurements has been proposed by Allan D.W. {[}4{]}
and used thoroughly by Witt T.J. in the case of voltage measurements
{[}5,6,7{]}.

In this work, a variety of TSA methods are being introduced and used
{[}8{]} in order to best analyze univariate time series of observations
collected in equally spaced time intervals. These data were obtained
in an attempt to characterize the performance of a newly commissioned
mass comparator for weighing in air. Eventually it will be used for
the determination of the mass of density artifacts such as Si spheres. 

The applied methods include both time and frequency domain techniques
such as the construction of Lag Plots, the use of the Auto-Correlation
function (ACF), the study of Power Spectral Density (PSD) {[}1,9{]}
and the two-sample or Allan Variance {[}10{]}. More specifically,
Lag Plots can reveal possible underlying data structures depending
on the observed data pattern. The analysis of the PSD can identify
possible system resonances and by employing noise power law models
{[}9{]} the type of noise present in the measured signal can, at least
qualitatively, be determined. The Allan Variance, which is an alternate
but equivalent time domain method, is more efficient in characterizing
the variance of autocorrelated measurements and the type of noise
present. Finally, the Autocorrelation Function gives the correlation
of consecutive measurements and the {}``memory span'' of the system. 

The analysis applied aims at finding autocorrelations in the time
series observations, characterizing the noise in the signal output
and providing a realistic estimate of the uncertainty of correlated
measurements {[}11{]}. The results from the above methods will be
compared to the power law models of three well-known types of noise
that exist in nature: flicker noise, white noise and random walk noise.
Even though the application of all the computational procedures described,
is performed on a series of measurements that comes from a precision
balance system, they can be applied to any electromechanical transducer
system equally well with similar results.

\section{Time Series Analysis Methods}

Time Series are an example of a stochastic or random process. Mathematically,
a stochastic process is an indexed collection of random variables
\{X$_{t}$: $t\in T$\}. When such a process is treated as a time
series, then the process is indexed by time, either discrete or continuous
{[}8{]}. We will specifically focus on discrete time univariate time
series analysis, by using the methods presented below:

\subsection{Stationary Processes}

For most of the time series encountered in several scientific fields,
the correlations between successive measurements have a deterministic
component (linear trend, periodic characteristics etc) and a stochastic
one. The factor that is in most cases responsible for a linear trend
in a set of measurements is the ambient temperature. Any deterministic
trend of that kind leads to non-stationary processes (non-stationary
mean or non-stationary variance) where the corresponding statistical
properties do not converge to any particular value. 

In this paper, the effects of these trends have been removed by the
application of the differencing method, transforming the non-stationary
time series to stationary ones. Considering the first differences
of the series \{X$_{t}$\} yields the same result as applying a high-pass
filter to the output signal:

\begin{equation}
\nabla X_{t}=X_{t}-X_{t-1}\end{equation}

Moreover, the application of the second differences filter, eliminates
any periodic characteristics of the time series:

\begin{equation}
\nabla^{2}X_{t}=X_{t}-2\cdot X_{t-1}+X_{t-2}\end{equation}

The purpose of the above transformation is the removal of any deterministic
attributes that would screen possible stochastic characteristics of
the system that produces the time series. The transformed series is
thereafter examined with time domain as well as frequency domain techniques
in order to reveal its stochastic character.

\subsection{Lag Plots}

The simplest test of extracting the correlation of two variables is
the scatter plot of the dependent variable as a function of the independent
one. The existence or not of a trend in such a scatter plot indicates
the correlation or not between the plotted variables. 

The same method is applicable in a time series of events where these
scatter plots are called lag plots {[}1{]}. The term \textquotedblleft{}lag\textquotedblright{}
defines a specific step in time. Therefore, a lag plot is a plot of
the observation X$_{t}$ versus the observation X$_{t-\tau}$, \textgreek{t}=lag
interval or time, that provides evidence of the possible autocorrelations
in a time series. 

The interpretation of the lag plot can answer to the questions of
data randomness, serial correlations, identification of outliers and
suggest a proper model that best describes the data set. According
to the underlying structure of the data set, the lag plot can identify
4 different case studies: White Noise, strong positive correlations
(Random Walk Noise), weak positive correlations (Pink Noise) and negative
correlations (Phase Noise).

\subsection{AutoCorrelation Function}

In the case of any two random variables X, Y the equation for the
variance of their sum is given by equation (3):

\begin{center}
\begin{equation}
var[X+Y]=var[X]+var[Y]+2cov[X,Y]\end{equation}

\par\end{center}

Generalizing equation (3) for the variance of the mean of N observations,
we get {[}3{]}

\begin{equation}
var[\bar{X]}=\frac{1}{N^{2}}\sum_{i=1}^{N}var(x_{i})+\frac{2}{N^{2}}\sum_{j=1}^{N}\sum_{k>j}^{N}R(x_{j},x_{k})\end{equation}
where the second term of the equation represents the autocovariance
between adjacent measurements of the univariate time series of length
N.

Alternatively we have

\begin{equation}
var[\bar{X]}=\frac{1}{N}var[X]\cdot\left(1+\frac{2}{N}\sum_{j=1}^{N}\sum_{k>j}^{N}\rho(x_{j},x_{k})\right)\end{equation}
where \textgreek{r}(x$_{j}$,x$_{k}$) represents the fraction of
autocovariance over variance and it is called autocorrelation. Considering
that autocovariance and thus autocorrelation of any two observations
depends only on their in-between time lag \textgreek{t} and not on
the ordering j, k we can simply derive the autocorrelation function
(ACF) \textgreek{r}(\textgreek{t}) and its estimator r(\textgreek{t})

\begin{equation}
r(\tau)=\frac{\sum_{i=1}^{N-\tau}(x(t_{i})-\overline{x})(x(t_{i}+\tau)-\overline{x})}{\sum_{i=1}^{N}(x(t_{i})-\overline{x})^{2}}\end{equation}

According to the shape of the autocorrelation function {[}12{]}, the
following cases can be recognized: (i) No specific shape, where the
white noise model can be applied and data fall within 95\% confidence
bands of the ACF plot (or correlogram). (ii) Exponential or alternating
positive and negative values and then decaying to zero, which is indicative
of an Autoregressive (AR) model. (iii) One or more spikes and the
rest zero, where a Moving Average (MA) model is appropriate. (iv)
Decay starting after a few lags, where a mixed ARMA model is recommended.
(v) No decay to zero, where the process is non-stationary.

\subsection{Power Spectral Density}

The Power Spectrum is a very popular method used in the characterization
of physical systems {[}9{]}. By simply applying a Discrete Fourier
Transform to the time series data, the real Re(X(f)) and imaginary
Im(X(f)) parts of the Fourier coefficients are derived. The periodogram
is a plot of the magnitude squared of the coefficients S(f) versus
frequency f, constituting the Power Spectral Density (PSD).

Moreover, the PSD is the Fourier Transform of the Autocorrrelation
Function \textgreek{r}(\textgreek{t}) and that relation, given by
equation (7), makes these techniques equivalent in the analysis of
time series:

\begin{equation}
S(f)=\intop_{-\infty}^{+\infty}\rho(\tau)\cdot exp(-2\pi if\tau)d\tau\end{equation}

By plotting S(f) \textendash{} f in a log-log diagram, the different
noise levels in the power spectrum can be estimated from the adjustment
of equation (8) to the log-log plot. The calculation of the spectral
exponent n, leads to the identification of white noise (n=0), flicker
noise (n=1), random walk noise (n=2) or any intermediate case 0<n<2
of colored noise.

\begin{equation}
S(f)=\frac{h_{n}}{f^{n}},\:0\leq n\leq2\end{equation}

\subsection{Allan Variance}

The Allan Variance or Two-sample Variance was first introduced by
David W. Allan for the evaluation of the stability of time and frequency
standards {[}10{]}. The main idea behind this method lies in the relation
of the dispersion that a set of measurements has, with the noise we
expect to find in the output signal. The most common measure of dispersion
is the classical variance, whose value decreases as the number of
data points included in the calculations, increases. 

Unfortunately this is only true in the case of truly random processes,
where the variance of the mean decreases with the number N of data
points

\begin{equation}
var[\overline{X}]=\frac{\sigma^{2}}{N}\end{equation}

In the case of autocorrelated data the above equation does not apply
since there exists a possibility that the estimated variance will
diverge as the number of data points increases {[}4{]}. So, we create
a high-pass filter by extracting each k+1 value from its previous
one k and thus we remove any possible trends, fast fluctuations or
other peculiar characteristics. The Allan variance is estimated at
time intervals \textgreek{t}=m\textgreek{t}$_{0}$, where \textgreek{t}$_{0}$
is a minimum sampling time and m is usually chosen to denote powers
of two. From the resulting plot of $\sigma(\tau)-\tau$ one can estimate
the cut-off value after which the inclusion of more data points does
not lead to the decrease of the variance, as classically indicated
by equation (9). In a log-log plot the Allan variance is proportional
to \textgreek{t}$^{\mu}$ and \textgreek{m} = -n-1, where n is the
spectral exponent appearing in equation (8).

\begin{equation}
\sigma_{y}^{2}(\tau)=\frac{1}{2(N-1)}\sum_{k=1}^{N-1}(y_{k+1}(\tau)-y_{k}(\tau))^{2}\end{equation}

In this paper lag plots and autocorrelation function are used for
the analysis of correlations in the time series while spectral analysis
and Allan variance are used to test time series for colored noise
and estimate realistic uncertainties. It should be noted that a normal
distribution of observations couldn\textquoteright{}t tell between
white noise and colored noise thus the mere use of histograms for
the acceptance or not of equation (9) is pointless.

\section{Experimental}

Data was obtained from characterization experiments of a newly commissioned
1kg/ 10\textgreek{m}g resolution mass comparator at the Hellenic Institute
of Metrology density laboratory. The principle function of this comparator
will be the measurement of mass in air of Si and ceramic density artifacts
of mass ca. 1kg in the form of spheres through comparative weighing.
Due to the limited space on its weighing pan as well as the weighing
procedure itself it was deemed necessary to also use its under-floor
weighing facility in order to perform the measurements. Thus the object
of which the mass is to be determined is placed in a ring cradle and
hung from the balance weighing cell via a ca. 80cm steel wire which
passes through a hole at the bottom of the balance. In order to have
stable weighing conditions, a plexi-glass draft shield of approximately
0,3 $m{}^{3}$ volume was used with a facility for installing temperature
and relative humidity sensors. A simple weight exchange mechanism
was used to alternately weigh the weight standard placed on the balance
pan and the test object hanging from the balance in an attempt to
minimize the disturbance of the weighing environment {[}13{]}. Since
weighing of non-conducting objects is prone to electrostatic charging
effects it was ensured that the relative humidity within the draft
shield was above 40\%. The air temperature was set at 22,5 $^{0}C$.
The data presented below resulted from a Si sphere of mass approximately
1kg hanging from the balance weighing cell. Sampling was performed
every 20 sec of the balance indication and of the environmental conditions.

\section{Results}

The following figures show the raw data transformed into a time series
of equal time intervals (figure 1), the variation of temperature (figure
2) and relative humidity (figure 3) over time.

For a time interval approximately equal to three hours, the time series
evolves in accordance with ambient temperature, a consequence of the
increase of air density and thus of buoyancy due to air. Afterwards,
it shows an increasing trend most probably due to balance drift. In
order to eliminate these deterministic affects, a first difference
filter is applied and the resulting series is thoroughly examined
for stochastic correlations and noise.

\noindent \begin{center}
\includegraphics[width=10cm,height=6cm,keepaspectratio]{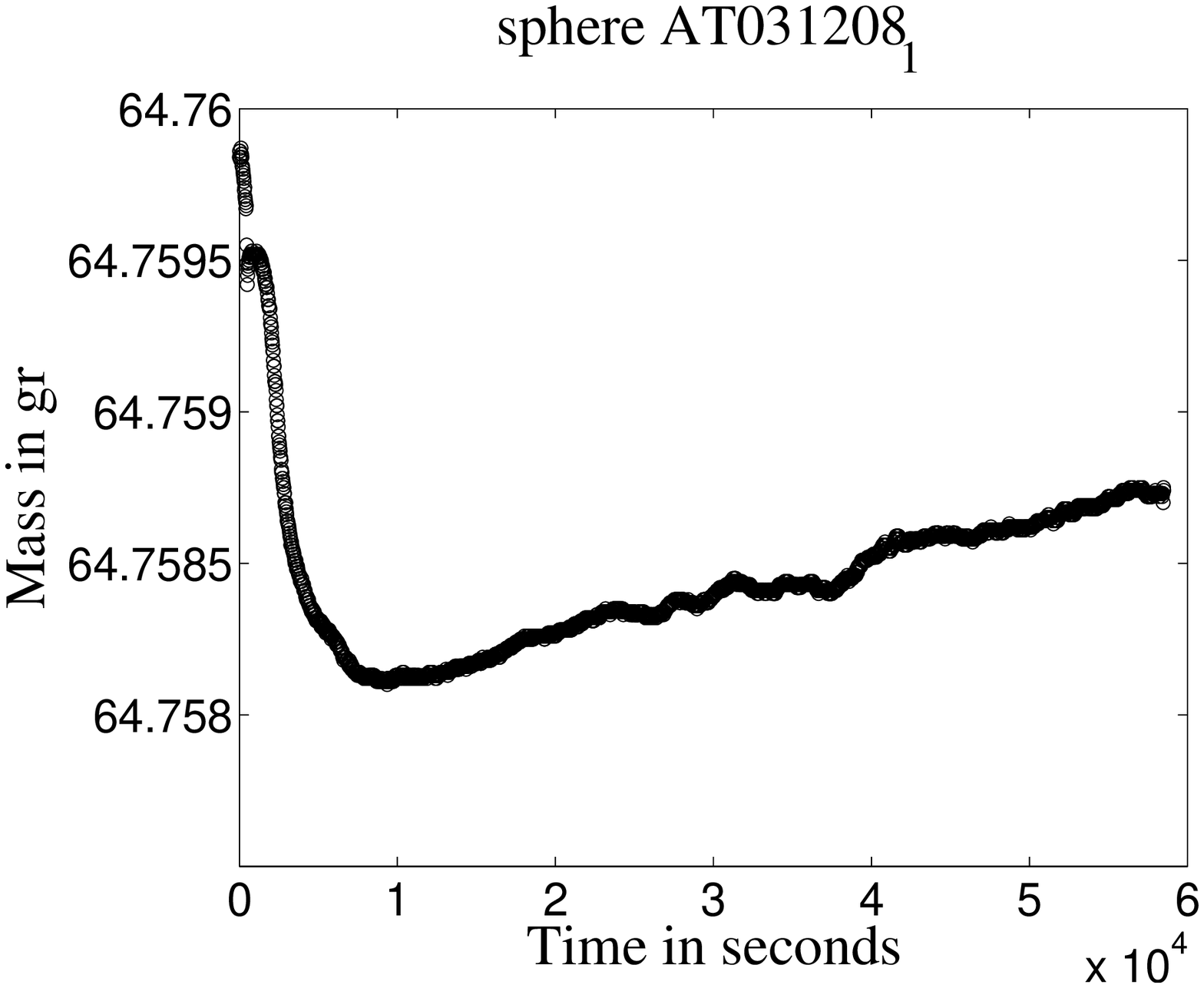}
\par\end{center}

\noindent \begin{center}
Figure 1. Time series of hanging Si sphere.
\par\end{center}

\noindent %
\begin{minipage}[t]{0.5\columnwidth}%
\begin{center}
\includegraphics[width=7cm,height=7cm,keepaspectratio]{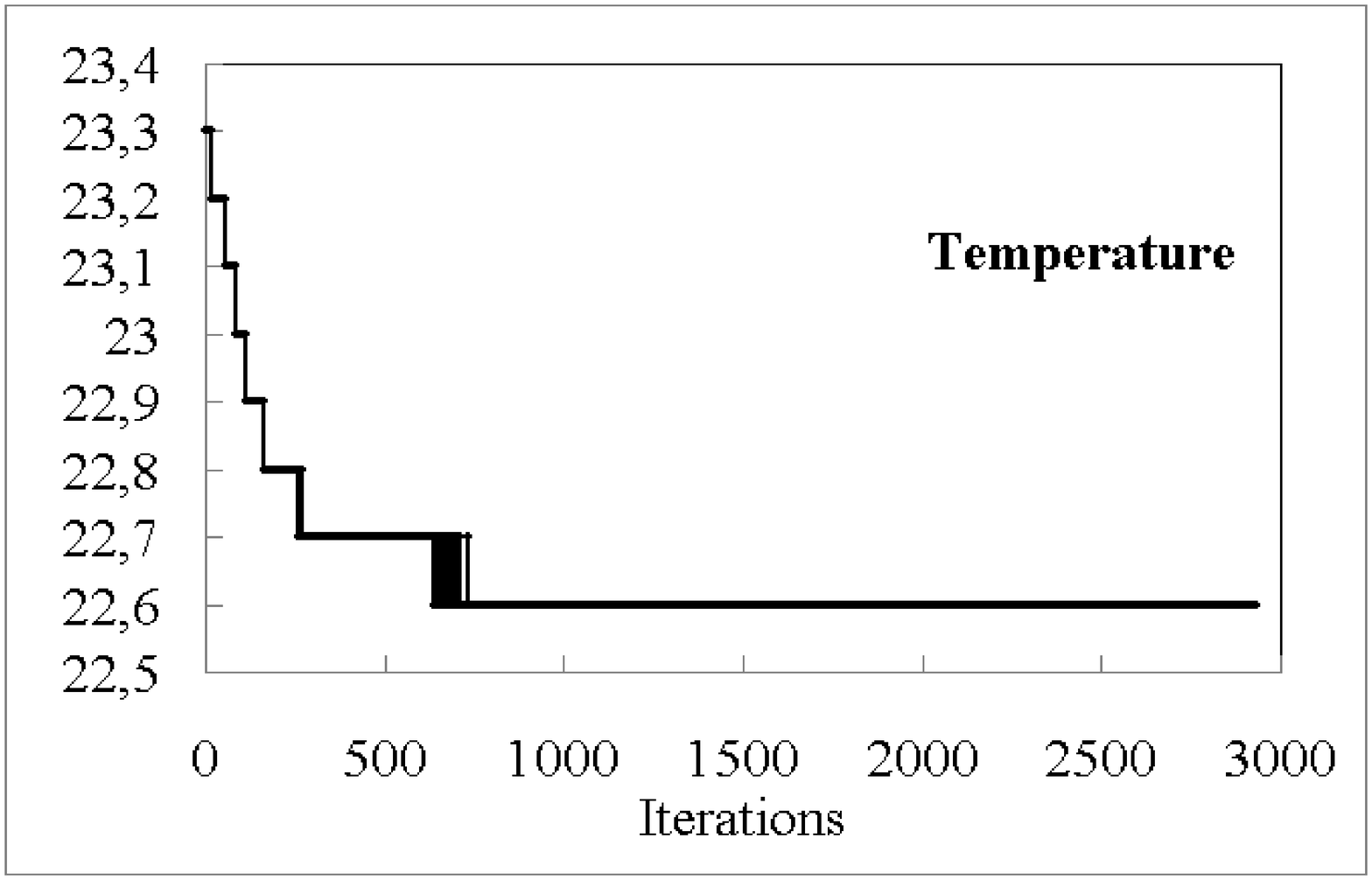}
\par\end{center}

\noindent \begin{center}
Figure 2. Ambient conditions, temperature.
\par\end{center}%
\end{minipage}\hfill{}%
\begin{minipage}[t]{0.5\columnwidth}%
\begin{center}
\includegraphics[width=7cm,height=7cm,keepaspectratio]{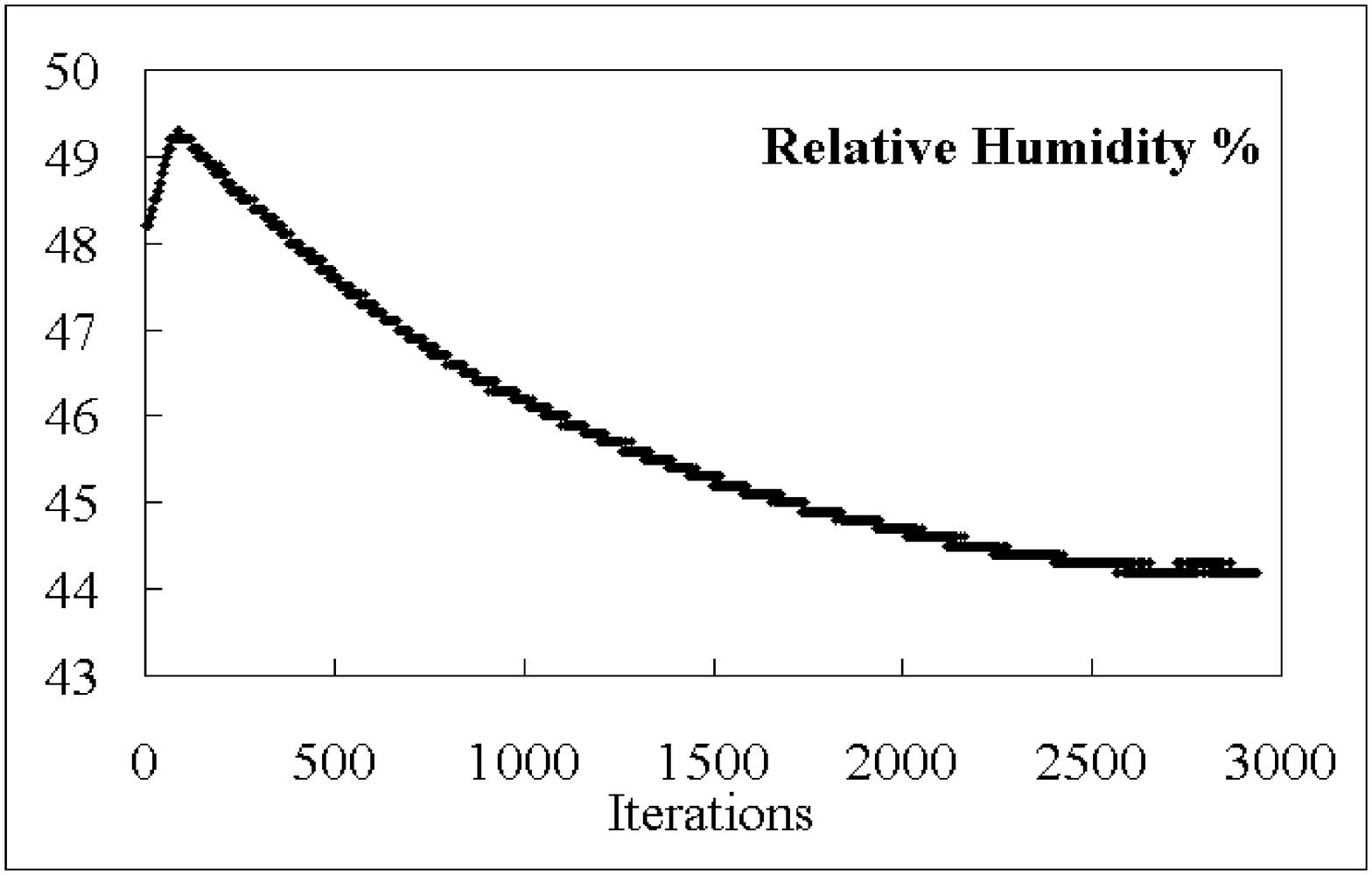}
\par\end{center}

\noindent \begin{center}
Figure 3. Ambient conditions, relative humidity.
\par\end{center}%
\end{minipage}

The lag plot of the differenced series in figure 4 is comprised of
a small number of points, due to the limited discritization available,
gathered around (0,0) with no particular structure that could reveal
any serial correlations.

\begin{center}
\includegraphics[width=8cm,height=8cm,keepaspectratio]{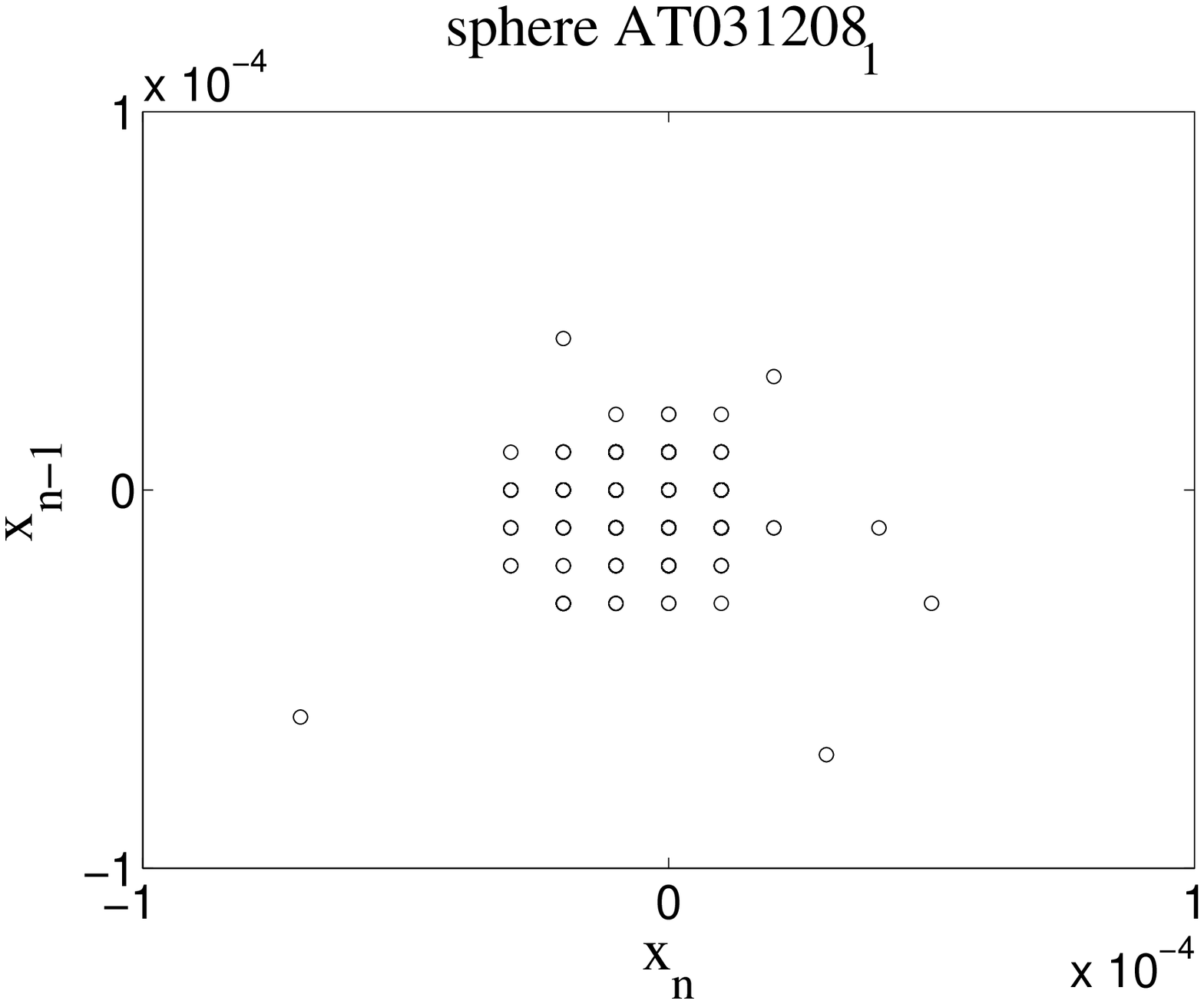}
\par\end{center}

\begin{center}
Figure 4. Lag Plot of the differenced time series, showing no structure.
\par\end{center}

$\vphantom{}$

The correlogram in figure 5 shows positive serial correlations in
the first lags and the sample ACF approaches zero very slowly (after
2000 sec), suggesting the inadequacy of the white noise model and
the use of classical variance. From approximately the 200th lag and
after, the data become independent falling well within the 95\% confidence
bands of $\pm\frac{2}{\sqrt{N}}$, implying that the white noise model
is now sufficient to describe the measurements. Consequently, two
different noise areas are also expected in the PSD diagram. The shape
of the ACF is exponentially decaying to zero so an AR model would
be appropriate to model the data {[}5{]}.

\begin{center}
\includegraphics[width=8cm,height=8cm,keepaspectratio]{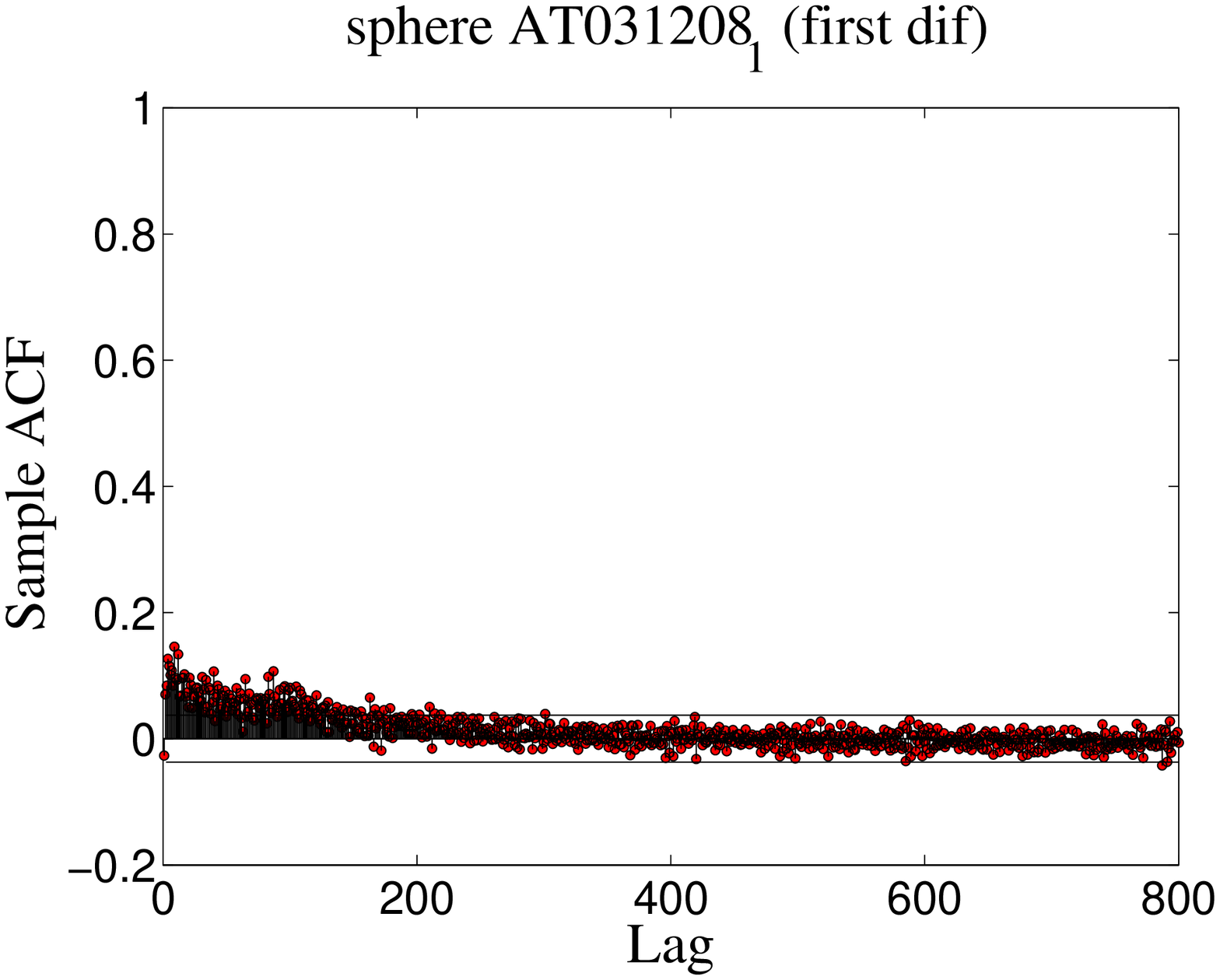}
\par\end{center}

\noindent Figure 5. The sample autocorrelation function of the first-differenced
data, showing the long-term correlations of the first lags. The horizontal
lines indicate the 95\% confidence levels for the white noise model.

$\vphantom{}$

The periodogram of the differenced data as well as the PSD (in log-log)
of the original and the differenced series are presented in figures
6 and 7. In figure 6, the presence of strong power intensity in lower
frequencies indicates an AR model with \textgreek{f}>0 to model the
data. Figure 7 shows a comparison between the original time series
before the removal of deterministic affects and the differenced series.
As expected, a Random-Walk Noise Model adequately describes the non-stationary
series and the slope of the regression line (red line) is close to
\textendash{}2. On the contrary, the differenced series exhibits a
more complex behavior. At lower frequencies, a line of slope close
to \textendash{}0.6 (blue line) can approximate the data but at higher
frequencies the white noise model of zero slope seems to be more appropriate
(green line). Thus, there exists a cut-off frequency at which a marked
change of slope occurs, revealing the two different noise areas of
flicker or pink noise and white noise {[}5{]}. This implies that after
about 500 sec the classical variance would be valid as a measure of
dispersion of the data. In addition, a suitable power law model would
be $S{}_{y}(f)=h_{0}f^{0}+h_{-1}f^{-1}$, where $h{}_{0}$ and $h{}_{-1}$
are the intensity coefficients. Table I summarizes the results for
the two time series.

\begin{center}
\includegraphics[width=8cm,height=8cm,keepaspectratio]{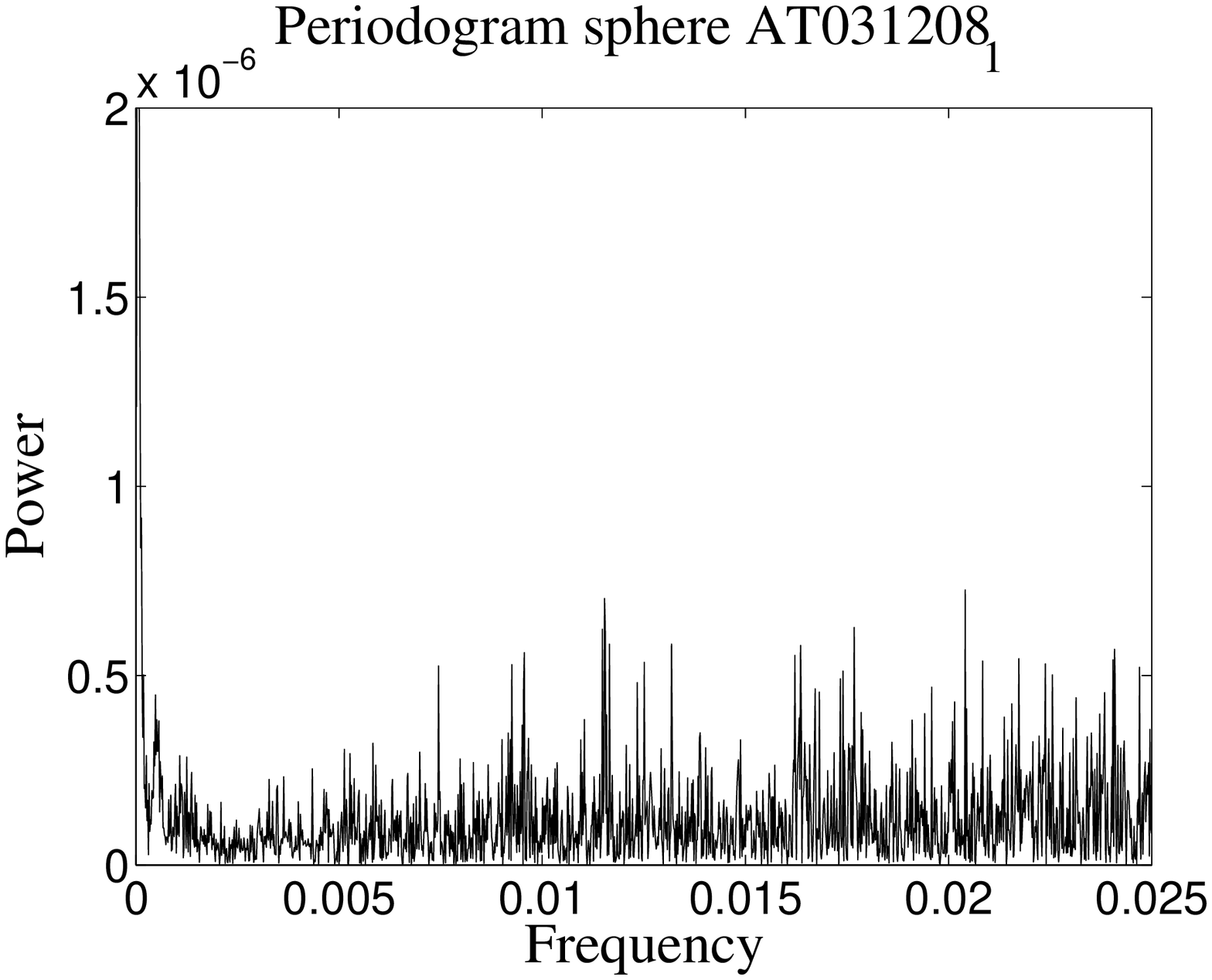}
\par\end{center}

\noindent Figure 6. The periodogram of the first-differenced data,
showing high intensity at low frequencies and white noise at the rest
of the spectrum.

\begin{center}
\includegraphics[width=8cm,height=8cm,keepaspectratio]{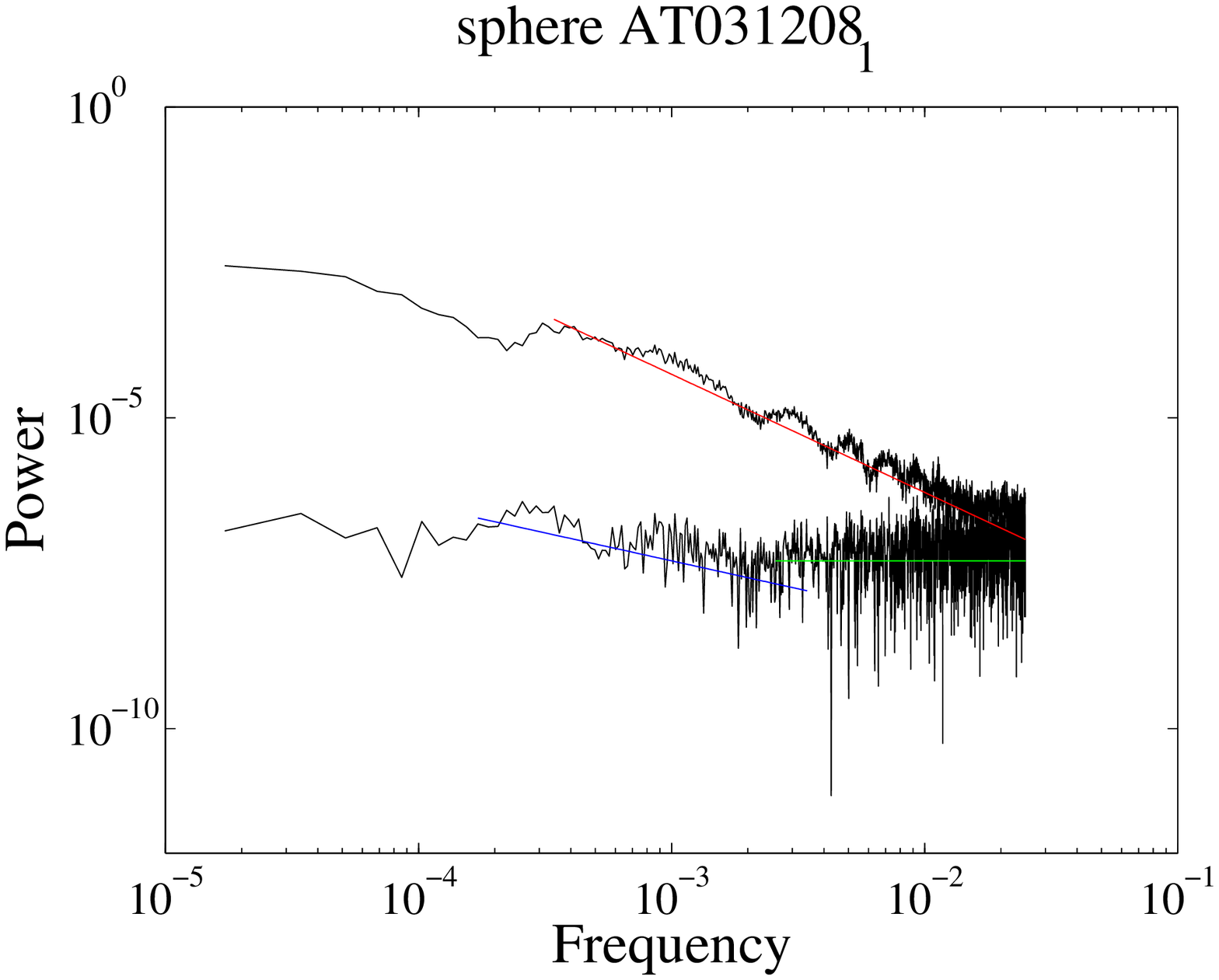}
\par\end{center}

\noindent Figure 7. The PSD of the original compared to the first-differenced
data. After eliminating the deterministic affects, the random walk
noise turns into a combination of flicker and white noise. The fitted
line is of the form y = \textendash{} n{*}x + logh, where y=log(Power)
and x=log(Frequency).

$\vphantom{}$

\pagebreak{}

\begin{center}
Table I.
\par\end{center}

\begin{center}
\begin{tabular}{|c|c|c|c|}
\hline 
 & Area 1 & Area 2 & Area 2 start\tabularnewline
\hline
\hline 
n (orig) & -1.755 & - & -\tabularnewline
\hline 
h (orig) & 7.45 $10{}^{-5}$ & - & -\tabularnewline
\hline 
n (firstdif) & -0.554 & +0.25 & 2 $10{}^{-3}$\tabularnewline
\hline 
h (firstdif) & 1.7 $10{}^{-4}$ & 1.36 $10{}^{-3}$ & 2 $10{}^{-3}$\tabularnewline
\hline
\end{tabular}
\par\end{center}

\begin{center}
$\vphantom{}$
\par\end{center}

From the above analysis, one concludes that the use of Allan Variance
could be more appropriate than Classical Variance, particularly during
the initial time duration {[}4,5{]}. The dependence of the Allan deviation
that is shown in figure 8 is in complete agreement with the PSD results,
indicating clearly the white noise and the flicker noise area. Notice
that the last three points, if included in the calculation, will lead
to an overestimation of the sample deviation. Indeed, the classical
standard deviation is equal to 6.53 10$^{-6}$ where the Allan deviation
gives a value of one less order of magnitude \textgreek{sv}(\textgreek{t})
= 7.34 10$^{-7}$ for \textgreek{t}=256. The last three data points
after $\tau$=256, are indicative of the flicker noise and thus are
not included in the estimation of the sample variance.

\begin{center}
\includegraphics[width=8cm,height=8cm,keepaspectratio]{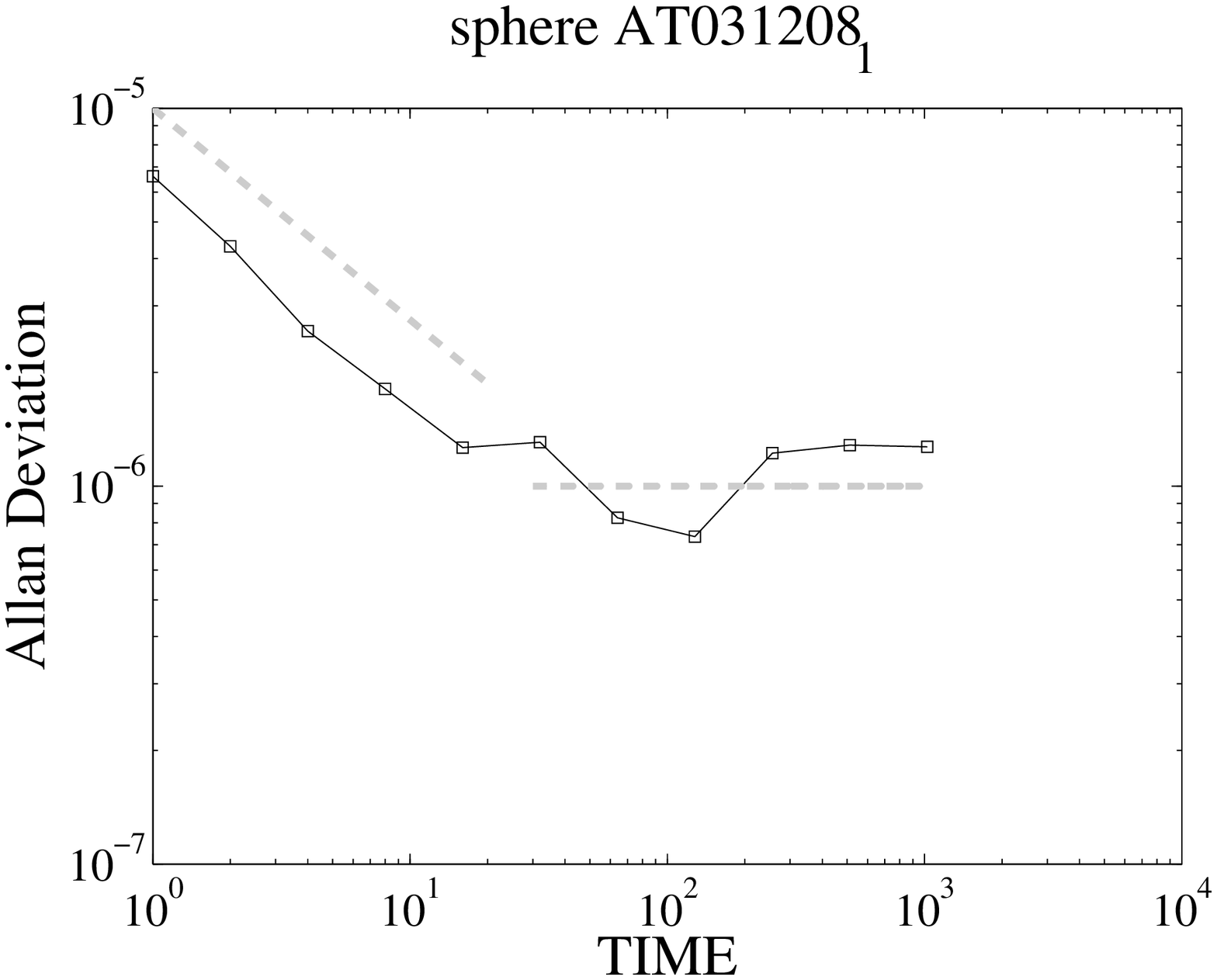}
\par\end{center}

\noindent \begin{center}
Figure 8. The Allan deviation of the first-differenced series, showing
white and flicker noise.
\par\end{center}

Table II gives the results from the fitting and the estimation of
exponent \textgreek{m}, as well as the calculation of the spectral
exponent from the relation n = -\textgreek{m}-1. White noise and flicker
noise areas are clearly indicated.

\begin{center}
Table II.
\par\end{center}

\begin{center}
\begin{tabular}{|c|c|c|}
\hline 
 & $\mu$ & n\tabularnewline
\hline
\hline 
Area 1 & -0.8942 & -0.1058\tabularnewline
\hline 
Area 2 & $0.027$ & -1.02\tabularnewline
\hline
\end{tabular}
\par\end{center}

\section{Conclusions}

The main purpose of this work is to point out the importance of considering
measurement results as time series and of using the time series analysis
methods to test for correlations. Only in the absence of correlation
the use of the expression $\sigma/\sqrt{N}$ is appropriate to characterize
the random uncertainty. Otherwise, the Allan variance is an alternate,
more appropriate tool to characterize the dispersion of a set of experimental
results. 

Lag Plots, Correlograms and Periodograms are graphic tools of EDA
analysis used to identify serial correlations in measurements and
estimate the type of the underlying stochastic noise. For the case
study of mass measurements we consider here, all the above techniques
give consistent results. 

Allan Variance and PSD analysis provide almost the same information
for a set of measurements but with the strict limitation of data collected
at regular time intervals. Also, the length of the time series must
be of a value of 2$^{k}$ otherwise not all points are included in
the Fourier Transform and Allan Variance calculations. 

Future work will mainly focus on the application of TSA methods to
other measurement circumstances, the comparison of the PSD with Bayesian
spectrum analysis methods, the possible use of a second difference
filter and the phase noise analysis.

\end{document}